\documentclass[twocolumn, aps,amsmath,amssymb,amsfonts, superscriptaddress]{revtex4}
\usepackage[german,american,english]{babel}
\usepackage{graphicx}
\usepackage{graphics}
\usepackage{dcolumn}
\usepackage{bm}
\usepackage{amssymb}
\usepackage{amsmath}
\usepackage{amsfonts}
\usepackage{epsfig}

\newcommand {\bra} [1] {\langle #1 |}
\newcommand {\ket} [1] {| #1 \rangle}
\newcommand {\bkt} [1] {\langle #1 \rangle}
\newcommand {\dbkt} [2] {\langle #1 | #2 \rangle}
\newcommand {\tbkt} [3] {\langle #1 | #2 | #3 \rangle}

\begin{document}
\title{Valley-based noise-resistant quantum computation using Si quantum dots}
\author{Dimitrie Culcer}
\affiliation{ICQD, Hefei National Laboratory for Physical Sciences at the Microscale, University of Science and Technology of China, Hefei,
Anhui 230026, China}
\author{A. L. Saraiva}
\affiliation{Instituto de F\'isica, Universidade Federal do Rio de Janeiro, Caixa Postal 68528, 21941-972 Rio de Janeiro, Brazil}
\author{Belita Koiller}
\affiliation{Instituto de F\'isica, Universidade Federal do Rio de Janeiro, Caixa Postal 68528, 21941-972 Rio de Janeiro, Brazil}
\author{Xuedong Hu}
\affiliation{Department of Physics, University at Buffalo, SUNY, Buffalo, NY 14260-1500}
\author{S.~Das Sarma}
\affiliation{Condensed Matter Theory Center, Department of Physics, University of Maryland, College Park MD20742-4111}
\begin{abstract}
We devise a platform for noise-resistant quantum computing using the valley degree of freedom of Si quantum dots. The qubit is encoded in two polarized (1,1) spin-triplet states with different valley compositions in a double quantum dot, with a Zeeman field enabling unambiguous initialization. A top gate gives a difference in the valley splitting between the dots, allowing controllable interdot tunneling between opposite valley eigenstates, which enables one-qubit rotations. Two-qubit operations rely on a stripline resonator, and readout on charge sensing. Sensitivity to charge and spin fluctuations is determined by intervalley processes and is greatly reduced as compared to conventional spin and charge qubits. We describe a valley echo for further noise suppression.
\end{abstract}
\maketitle

Quantum computing (QC) requires accurate control of the two states in a quantum bit (qubit), which entails long coherence times and ability to rotate a single qubit and entangle neighboring qubits. Solid state spin systems are an obvious choice for scalable QC, with Si known for its outstanding spin coherence times \cite{Feher_PR59, Abe_PRB04, Tyryshkin_JPC06, Hanson_RMP07}, thanks to lack of piezoelectric electron-phonon coupling \cite{Prada_PRB08}, weak spin-orbit interaction \cite{Wilamowski_Si/SiGeQW_Rashba_PRB02, Tahan_PRB05} and most importantly, nuclear-spin free isotopes, enabling isotopic purification to remove hyperfine coupling \cite{Witzel_AHF_PRB07}. Recently Si quantum dots (QDs) have been at the forefront of QC research, with theories of spin relaxation in Si QDs further justifying this choice \cite{Wang_SiQD_ST_Relax_PRB10, Raith_SiQD_1e_SpinRelax_PRB11}. Experimental progress in Si QC has been made using QDs in Si/SiO$_2$ \cite{Liu_PRB08, Xiao_MOS_SpinRelax_PRL10, Nordberg_APL09, Ferrus_09}, Si/SiGe \cite{MarcusGroup_NatureNano07, Shaji_NP08}, and donor-based architectures \cite{Stegner_NP06, Kuljanishvili_NP08, Lansbergen_NP08, Fuhrer_NanoLett09, Lim_SingleElectron_APL09, Morello_1glShot_Nature10, Lai_SiDQD_SpinBlock_10}.

The two main obstacles to Si QC are dephasing due to charge noise and the valley degree of freedom. In layered structures only the two low-lying valleys perpendicular to the interface are relevant, and the interface potential gives a valley-orbit coupling $\Delta$, which determines the two-spin exchange \cite{Hada_JJAP04}. The study of Si valley physics has intensified, in experiment \cite{Lai_PRB06, Takashina_PRL06, Goswami_NP07, Xiao_Valley_APL10, Borselli_Valley_APL11, Lim_SiQD_SpinFill_NT11} and theory \cite{Boykin_APL04, Friesen_PRB04, Nestoklon_PRB06, Boykin_PRB08, Srinivasan_APL08, Saraiva_PRB09, Friesen_PRB10}.

Here we propose a new QC platform in Si QDs using the valley degree of freedom to overcome noise. It exploits the closeness in energy of two polarized spin-triplet states with different valley compositions. It employs available technology in Si QD in MOSFET and donor architectures, offering all-electrical control and a fundamental perspective on valley manipulation \footnote{The principle differs from A.~Rycerz \textit{et al.}, Nat.\ Phys.\ \textbf{3}, 172 (2007) and P.~Recher \textit{et al.}, Phys.\ Rev.\ B \textbf{76}, 235404 (2007), which control the valley splitting in a \textit{single} graphene ribbon and ring respectively.
It also differs from Smelianskiy \textit{et al.}, Phys.\ Rev.\ B \textbf{72}, 081304 (2005), which use Li donors in Si.}. 

We take an inversion layer of thickness $L$, with a top gate setting the potential at the top Si/barrier interface to $V_T$ and a back gate inducing $V_B$ to the back interface. The electric field inside the inversion layer is $F=(V_B-V_T)/L$. 
In a single QD defined in this environment, an electron experiences the potential
\begin{equation}
V_D = \frac{\hbar^2}{2m^*a^2} \, \bigg[ \frac{(x - x_D)^2 + y^2}{a^2} \bigg] +  e F z + U_0 \, \theta(z) + e V_T,
\end{equation}
with the origin at the top interface. The dot is located at $(x_D, 0, 0)$, with a radius $a$, $m^*$ is the Si in-plane effective mass and $U_0 \, \theta(z)$ the interface potential with $\theta(z)$ the Heaviside function. The effective-mass ground state wave function is $D_\xi (x, y, z) = \phi_D(x,y) \, \psi (z) \, u_\xi ({\bm r})\, e^{ik_\xi z}$. The lateral envelope function $\phi_D(x,y)$ is a 2D Gaussian, and $\psi (z)$ is the solution to the triangular well imposed by the interface and the potential difference in the $z$ direction. The Bloch functions are $u_\xi ({\bm r})\, e^{ik_\xi z}$, where the valley index $\xi = \{ z, \bar{z} \}$, and $k_\xi = \pm k_0$, with $k_0 = 0.85 (2\pi/a_{Si})$, and $a_{Si} \! = 5.43 {\rm\AA}$ the Si lattice constant. In the basis $\{ D_\xi \}$ the single-dot Hamiltonian reads
\begin{equation}\label{H1e1d}
H_D = \varepsilon_D + \begin{pmatrix} 0 & \Delta_D \cr \Delta_D^* & 0 \end{pmatrix},
\end{equation}
with confinement energy $\varepsilon_D$ and valley-orbit coupling
\begin{equation}
\Delta_D = \tbkt{D_z}{U_0 \, \theta(z) + e F z}{D_{\bar{z}}}.
\end{equation}

The top and back gates independently control the diagonal confinement energy (electron density) and the off-diagonal valley-orbit coupling (vertical field.) Gate cross-talk can be circumvented using gate compensation techniques \cite{Morello_1glShot_Nature10}. Experimentally, the lever arm of the top and back gates must be obtained in order to determine how $V_T$ and $V_B$ control $\varepsilon_D$ and $\Delta_D$. The functional form of $\varepsilon_D (V_T,V_B)$ and $\Delta_D(V_T,V_B)$ may be obtained from theory \cite{Saraiva_PRB09}. 
The quasi-triangular potential in the $z$ direction is not exactly solvable. Nevertheless, it is a good approximation to adopt the exact expression for an infinite barrier $E_z=(\hbar^2/2m_z (1.13805 \pi e F)^2)^{1/3}$.~\cite{Saraiva_LargeVOC_PRB11} The total energy is $\epsilon_D= \hbar^2/m^* a^2+E_z+eV_T$. With $E_z$ fixed, $V_T$ shifts the diagonal terms in Eq.~(\ref{H1e1d}). Such control could also be realized in the setup of Ref.~\onlinecite{Lim_SiQD_SpinFill_NT11}, where Al-Al$_2$O$_3$ multi-gate stacks provide excellent tunability of the vertical electrical field and inter-dot coupling.

\begin{figure}[tbp]
\includegraphics[width=\columnwidth]{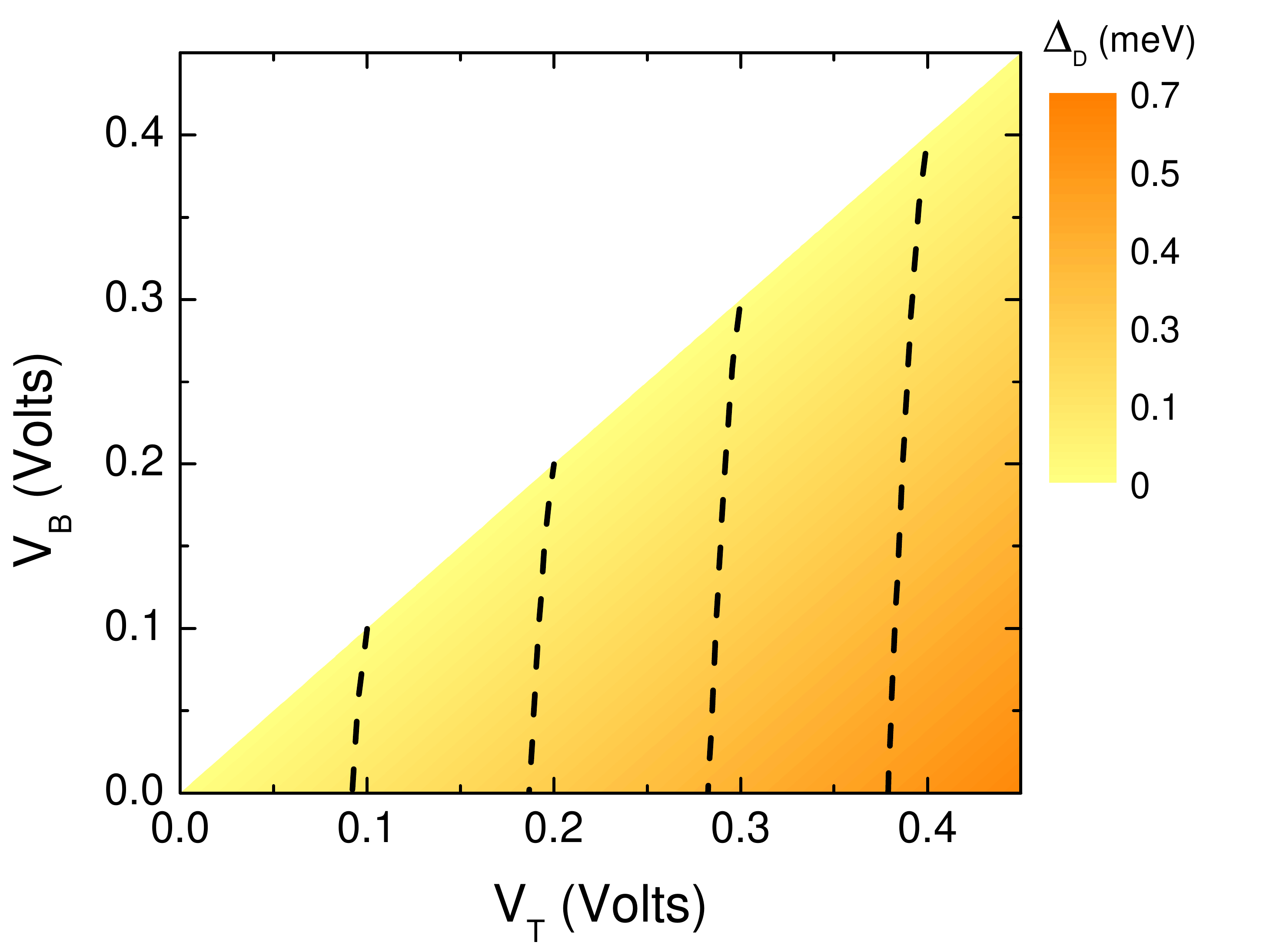}
\caption{Color plot of $|\Delta_D|$ as a function of the gate voltages. The dashed lines refer to paths of constant $\epsilon_D$.  If the qubit
is manipulated along these lines, the back gate voltage compensates the change in orbital energy induced by the top gate voltage. We
particularize our results to $V_T>V_B$, so that the electron is at the top interface at all times.} \label{fig:control}
\end{figure}

The valley-orbit coupling $\Delta_D = |\Delta_D| \, e^{-i\phi_D}$ is complex. The eigenstates of $H_D$ (\textit{valley eigenstates}) are
\begin{equation}\label{Dpm}
\ket{D_\pm} = (1/\sqrt{2}) \, (\ket{D_z} \pm e^{i\phi_D} \ket{D_{\bar{z}}}),
\end{equation}
with energies $\varepsilon_\pm$. The dominant contribution to $\Delta_D$ comes from $U_0$. The direct contribution from $F_D$ is smaller \cite{Saraiva_PRB09}, yet $F_D$ controls $\Delta_D$ by manipulating the electronic density at the interface, and $|\Delta_D|$ can be increased by an order of magnitude. For Si/SiO$_2$ barriers $|\Delta_D| = [(27.4\pm 0.02) F]$ meV, with $F$ in V/nm \cite{Saraiva_LargeVOC_PRB11}. Figure~\ref{fig:control} shows the way $V_T, V_B$ control $|\Delta_D|, \epsilon_D$ separately.

We focus on a double quantum dot (DQD), with the left dot located at $x_L = -x_0$ and the right dot at $x_R=x_0$. The valley eigenstates $\ket{D_\pm}$ are identical in both dots if the interface is sharp along the growth direction and flat perpendicular to it, or if interface roughness is correlated over distances much shorter than the size of the QD \cite{Culcer_Roughness_PRB10}. We assume the top and back gates can be adjusted independently for the L, R dots, so the electric field $F$ has different values for the two dots, $F_L$ and $F_R$ respectively, while their electrochemical potential is constant. The DQD confinement potential is
\begin{equation}
\arraycolsep 0.3 ex
\begin{array}{rl}
\displaystyle V_{DQD} = & \displaystyle \frac{\hbar^2}{2m^*a^2} \, \bigg\{\mathrm{Min} \, [(x-x_0)^2, (x+x_0)^2] + y^2\bigg\} \\ [3ex]
+ & \displaystyle eEx + eF_Lz + eF_Rz + U_0 \, \theta(z).
\end{array}
\end{equation}
Neglecting interdot couplings due to the interface potential (of order $l\Delta_D/\varepsilon_D$, with $l = \dbkt{L_\xi}{R_\xi} \ll 1$), $\Delta_D$ is a constant across the two dots and we diagonalize $H_D$ separately for $D = L, R$, obtaining $\Delta_L$, $\Delta_R$ and their corresponding valley eigenstates. These are orthogonalized (see Supplement), yielding the states $\ket{\tilde{D}_\pm}$. We use only $\ket{\tilde{D}_\pm}$ in the remainder of this work. Initially we take $\Delta_L = \Delta_R$, thus $\ket{\tilde{D}_\pm}$ are identical for $L, R$, and interdot tunneling preverves the valley eigen-index (i.e. between $\ket{\tilde{L}_\pm}$ and $\ket{\tilde{R}_\pm}$, but not between $\ket{\tilde{L}_\pm}$ and $\ket{\tilde{R}_\mp}$.)

We seek a two-dimensional subspace of the full DQD Hilbert space that consists of two states with different valley compositions that can be coupled controllably. The full Hilbert space includes spin and valley degrees of freedom. A high enough magnetic field ($\approx$ 1T) lifts the spin degeneracy, and only the lowest-energy $\downarrow\downarrow$ polarized triplet states are energetically accessible, thus below it is understood that $\tilde{T}^{LR}_{\pm\pm} \equiv \tilde{T}^{LR, \downarrow\downarrow}_{\pm\pm}$ etc. 
\begin{equation}
\arraycolsep 0.3 ex
\begin{array}{rl}
\displaystyle \tilde{T}^{LR}_{\pm\pm} = & \displaystyle (1/\sqrt{2}) \, \big( \tilde{L}_\pm^{(1)} \tilde{R}_\pm^{(2)} - \tilde{L}_\pm^{(2)}
\tilde{R}_\pm^{(1)} \big) \\ [1ex]
\displaystyle \tilde{T}^{LR}_{\pm\mp} = & \displaystyle (1/\sqrt{2}) \, \big( \tilde{L}_\pm^{(1)} \tilde{R}_\mp^{(2)} - \tilde{L}_\pm^{(2)}
\tilde{R}_\mp^{(1)} \big)
\\ [1ex]
\displaystyle \tilde{T}^{RR}_{+-} = & \displaystyle (1/\sqrt{2}) \, \big( \tilde{R}_+^{(1)} \tilde{R}_-^{(2)} - \tilde{R}_+^{(2)}
\tilde{R}_-^{(1)} \big).
\end{array}
\end{equation}
The hopping integral $\tilde{t} = \tbkt{\tilde{L}_\xi}{H_0}{\tilde{R}_\xi} + \tbkt{\tilde{L}_\xi \tilde{L}_\xi}{V_{ee}}{\tilde{L}_\xi \tilde{R}_\xi}$, the exchange integral $\tilde{j}
= \tbkt{\tilde{L}_\xi^{(1)} \tilde{R}_\xi^{(2)}}{V_{ee}}{\tilde{L}_\xi^{(2)} \tilde{R}_\xi^{(1)}} = \tbkt{\tilde{L}_\xi^{(1)}
\tilde{R}_{-\xi}^{(2)}}{V_{ee}}{\tilde{L}_{-\xi}^{(2)}  \tilde{R}_\xi^{(1)}}$, and $\tilde{\delta} = (\tilde{\varepsilon}_L - \tilde{\varepsilon}_R) - (\tilde{u} - \tilde{k})$ is an effective two-particle detuning between dots, with
$\tilde{u} = \tbkt{\tilde{D}_\xi^{(1)} \tilde{D}_\xi^{(2)}}{V_{ee}}{\tilde{D}_\xi^{(1)}  \tilde{D}_\xi^{(2)}} = \tbkt{\tilde{D}_\xi^{(1)}
\tilde{D}_{-\xi}^{(2)}}{V_{ee}}{\tilde{D}_\xi^{(1)} \tilde{D}_{-\xi}^{(2)}}$ the on-site Coulomb interaction and
$\tilde{k}=\tbkt{\tilde{L}_\xi^{(1)} \tilde{R}_\xi^{(2)}}{V_{ee}}{\tilde{L}_\xi^{(1)}\tilde{R}_\xi^{(2)}}$ the direct Coulomb interaction
between electrons on the two dots. In the basis $\{\tilde{T}^{LR}_{- -}, \tilde{T}^{LR}_{++}, \tilde{T}^{LR}_{+-},
\tilde{T}^{LR}_{-+}, \tilde{T}^{RR}_{+-}\}$ the effective Hamiltonian is 
\begin{equation}\label{H_T}
\arraycolsep 0.3 ex
\begin{array}{rl}
\displaystyle \tilde{H}_T = & \displaystyle \begin{pmatrix}
- |\tilde{\Delta}|_{tot} - \tilde{j} & 0 & 0 & 0 & 0 \cr
0 & |\tilde{\Delta}|_{tot} - \tilde{j} & 0 & 0 & 0 \cr
0 & 0 & |\tilde{\Delta}|_E & -\tilde{j} & \tilde{t} \cr
0 & 0 &  -\tilde{j} & - |\tilde{\Delta}|_E & - \tilde{t} \cr
0 & 0 & \tilde{t} & - \tilde{t} & - \tilde{\delta}
\end{pmatrix},
\end{array}
\end{equation}
with $|\tilde{\Delta}|_{tot} = |\tilde{\Delta}_L| + |\tilde{\Delta}_R|$ and $|\tilde{\Delta}|_E = |\tilde{\Delta}_L| - |\tilde{\Delta}_R|$. In the absence of interdot tunneling \textit{between opposite valley eigenstates}, the states $\tilde{T}^{LR}_{- -}$ and $\tilde{T}^{LR}_{++}$ are decoupled from the other three states, thus we do not include them in the rest of our analysis. The relevant eigenstates are
\begin{equation}
\begin{array}{rl}
\displaystyle \tilde{T}^<_{+-} = & \displaystyle \big(\varepsilon^<_0/\sqrt{\varepsilon^{<2}_0 + 2\tilde{t}^2}\big) \,
[ (\tilde{t}/\varepsilon^<_0) \, (\tilde{T}^{LR}_{+-} - \tilde{T}^{LR}_{-+}) + \tilde{T}^{RR}_{+-}] \\ [1ex]
\displaystyle \tilde{T}^{sym}_{+-} = & \displaystyle (1/\sqrt{2}) \, ( \tilde{T}^{LR}_{+-} + \tilde{T}^{LR}_{-+}) \\ [1ex]
\displaystyle \varepsilon^{<}_0 = & \displaystyle (1/2) \, [ -\tilde{\delta} + \tilde{j} - \sqrt{(\tilde{\delta} + \tilde{j})^2 +
8\tilde{t}^2} ].
\end{array}
\end{equation}
The qubit is defined by the states $\tilde{T}^<_{+-}$ and $\tilde{T}^{sym}_{+-}$. Restricting our attention to the qubit subspace
\begin{equation}
\arraycolsep 0.3 ex
\begin{array}{rl}
\displaystyle \tilde{H}_{Qubit} = & \displaystyle \begin{pmatrix}
-j & |\tilde{\Delta}|_E \cr
 |\tilde{\Delta}|_E & \tilde{\varepsilon}_0^<
\end{pmatrix} = -\tilde{j} -\frac{\hbar \omega_z}{2}+ \frac{\hbar}{2}
\begin{pmatrix}
\omega_z & \Omega \cr
\Omega & -\omega_z.
\end{pmatrix}
\end{array}
\end{equation}
Here $\hbar\omega_z = -\tilde{\varepsilon}_0^< -j = - \tilde{E}_{HM}$, the Hund-Mulliken energy that in one-valley systems denotes the singlet-triplet energy difference. In multivalley systems, due to the additional degree of freedom, $\tilde{E}_{HM}$ is not necessarily related to the spin part of the wave functions \cite{Culcer_PRB10}.

\begin{figure}[tbp]
\includegraphics[width=\columnwidth]{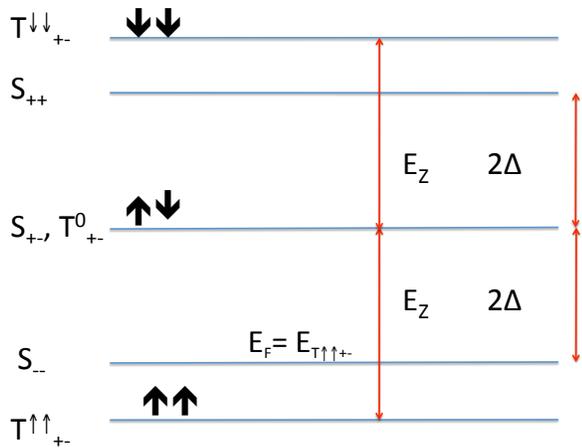}
\caption{Two-electron levels in a Si QD. For $E_Z > 2\Delta$, $T^{\downarrow\downarrow}_{+-}$ becomes the ground state and is initialized unambiguously.} \label{1e2d Levels}
\end{figure}

Referring to Fig.~\ref{1e2d Levels}, two electrons can be unambiguously initialized into $\tilde{T}^{RR}_{+-}$ when $\tilde{\delta}$ is raised and $\tilde{\varepsilon}_R \ll \tilde{\varepsilon}_L$. Then $\tilde{\delta}$ is swept to the (1,1) regime, with $\tilde{T}^{RR}_{+-}$ evolving into $\tilde{T}^{<}_{+-}$. This setup allows readout by charge sensing using a quantum point contact \cite{Petta_Science05}. Raising $\tilde{\delta}$ again, the two qubit states have different occupancies, realizing a \textit{valley blockade}. Readout of the charge state of one QD (e.g. R) automatically yields the qubit state.

A workable quantum computer requires tunable $x$- and $z$- rotations, i.e. the $\sigma_x$ and $\sigma_z$ gates.  A gate electric field $\parallel\hat{\bm z}$ generates the $\sigma_x$ gate. In the far detuned regime the states $\tilde{T}^{LR, \downarrow\downarrow}_{sym}$ and $\tilde{T}^{<, \downarrow\downarrow}_{+-}$ are effectively degenerate.  Ramping up $F_L$ induces a small difference $\tilde{\Delta}_E$ in the valley splitting on the left dot. With an electric field difference of $10^{-3}$ V/nm, $\tilde{\Delta}_E \approx 27.4 \mu$eV, and the gating time for a 2$\pi$ rotation around the $X$ axis is 0.15 ns. For $\sigma_z$ rotations $\tilde{E}_{HM}$ can be altered by changing $\tilde{\delta}$, providing a $\sigma_z$ gate of arbitrary strength. Thus, with $|\tilde{\Delta}_E| \gg \tilde{E}_{HM}$ or $\tilde{E}_{HM} \gg |\tilde{\Delta}_E|$, a $\sigma_x$ or a $\sigma_z$ gate can be obtained.

The $\sigma_x$ and $\sigma_z$ gates also allow a valley-echo experiment. Both $\sigma_x$ and $\sigma_z$ rotations can be gated quickly, providing a fast two-axis control, in contrast to e.g. singlet-triplet qubits, where one axis is the fixed Overhauser field gradient. With the available fast gates, advanced dynamical decoupling techniques \cite{Witzel_PRL07, Yao_PRL07} can restore coherence. If $\tilde{\Delta}_E < \tilde{E}_{HM}$, $X$ rotations may be implemented through valley resonance (cf. Ref.~\onlinecite{Palyi_PRL11}.) An alternating $\tilde{\Delta}_E(t)$ at a frequency resonant with $\omega_z$ enables Rabi-like flips of the valley qubit.

Initialization requires only a magnetic field, while in the far detuned regime, where operations are performed, the magnitude of $\tilde{\Delta}$ is irrelevant. Coherent rotations can be demonstrated without precise knowledge of $\tilde{\Delta}$, even for $\tilde{\Delta} \le k_BT$. Knowledge of $\tilde{\Delta}$ is required for \textit{controlled} rotations, whose characteristic time scale and coherence time may be estimated by Ramsey interferometry.

The universal set of quantum gates is completed using a superconducting transmission line resonator to provide entanglement. Superconducting resonators have exceptionally small mode volumes for strong couplings to qubits, and very high quality factors. They can mediate interactions between singlet-triplet qubits \cite{BurkardImam_DQD_QED_PRB06, Taylor_DQD_CavityQED_06} and can couple valley qubits. An array of DQDs can each be coupled to a superconducting transmission line resonator, with only the right dot coupled out of each DQD. The (quantized) resonator voltage is an addition to the detuning, and the effective interaction is
\begin{equation}
H_{res} = g \, (\hat{a} + \hat{a}^\dag) \, \ket{\tilde{T}^{RR}_{+-}}\bra{\tilde{T}^{RR}_{+-}},
\end{equation}
where $\hat{a}$ is the annihilation operator for the lowest energy mode of the resonator. $H_{res}$ is mapped onto a Jaynes-Cummings Hamiltonian. For Si shallow donor electrons coupling to Si cavities, $g$ has been estimated at 30 MHz \cite{Abanto_Si_Cvt_QC_PRB10}, and for QDs we expect an improvement of an order of magnitude. Type-II superconducting alloy striplines have high $B_{c2}$: NbTi, which is already used, has $B_{c2}$=15 T, and can be operated at the magnetic fields required here to freeze out the spin degree of freedom. 

Valley qubits have good coherence properties. Firstly, elaxation mediated by a generic phonon potential $V_{phn}$ is quantified by $T_1$, and strongly suppressed, because
\begin{equation}
\frac{1}{T_1} = \frac{2\pi}{\hbar} \sum_{\bm k} |\tbkt{\tilde{D}_+}{V_{phn}}{\tilde{D}_-}|^2 \delta(\varepsilon_+ - \varepsilon_- - \hbar \omega)
\end{equation}
requires phonons with energies $\hbar\omega \approx 2\tilde{\Delta}$ and ${\bm k}$ of the order of ${\bm k}_0$. Optical phonons bridge the difference in ${\bm k}$ but have energies $\gg 2\tilde{\Delta}$, whereas acoustic phonons with energies $\approx 2\tilde{\Delta}$ have wave vectors $\ll {\bm k}_0$ \cite{Yu_Cardona}. For the same reason, the qubit does not relax to the lower-lying $\tilde{T}^{LR}_{--}$ states with orthogonal valley composition.

Valley qubits are immune to charge noise due to dangling bonds (DBs), which hamper conventional charge and spin qubits \cite{Culcer_APL09}. Since a large wave vector $2k_0$ separates the $z$ and $\bar{z}$ valleys, intervalley matrix elements are only appreciable for real-space interactions sharp in the $\hat{\bm z}$-direction. For interactions non-singular in $z$, matrix elements between $\ket{\tilde{D}_z}$ and $\ket{\tilde{D}_{\bar{z}}}$, likewise between $\ket{\tilde{D}_+}$ and $\ket{\tilde{D}_-}$, are suppressed. The Coulomb potential $v(x,y)$ of a DB in the QD plane has intervalley matrix element $\bkt{v} = \tbkt{\tilde{D}_z}{v(x,y)}{\tilde{D}_{\bar{z}}} = 0$, since the overlap of valley states vanishes (see Supplement.) For a DB not in the plane, the intervalley matrix element of $v(x,y)$ is still suppressed, since quasi-2D screened Coulomb potentials remain long range. Thus the $\sigma_x$ gate is immune to noise. 

Noise in the $\sigma_z$ gate could lead to pure dephasing \cite{Culcer_APL09}. Variations in $\tilde{E}_{HM} $ are given by $\Delta \tilde{E}_{HM} = (\partial \tilde{E}_{HM}/\partial \tilde{t}) \, \Delta \tilde{t} +
(\partial\tilde{E}_{HM}/\partial\tilde{\delta}) \, \Delta \tilde{\delta}$. In the far-detuned regime, where the qubit is operated, there is
little variation in $\tilde{t}$ \cite{Culcer_APL09}. With the $\sigma_x$ gate off, $\tilde{H}_{Qubit} = (\tilde{E}_{HM} + \Delta \tilde{E}_{HM}) \, \sigma_z$ and consider the decay of the off-diagonal element $\rho_{12}$ of the density matrix. Random telegraph noise causes $\rho_{12} \propto e^{-t/\tau} \,  \cos \eta t $, with $\eta \approx \Delta \tilde{E}_{HM}/\hbar$ and $\tau$ the switching time of the fluctuator, which can reach ms \cite{Culcer_APL09}. For $1/f$ noise, $\rho_{12} \propto e^{-\chi(t)}$, where
\begin{equation}
\chi (t) = \frac{1}{2\hbar^2}\, \bigg( \frac{d \tilde{E}_{HM}}{d \tilde{\delta}} \bigg)^2 \int_{\omega_0}^{\infty}d\omega \,
S_{\tilde{\delta}}(\omega) \, \bigg(\frac{\sin\omega t/2}{\omega/2}\bigg)^2 \,.
\end{equation}
where $S_{\tilde{\delta}}(\omega)$ is the spectral density of fluctuations in $\tilde{\delta}$ and $\omega_0$ a cut-off (the inverse measurement time.) Since the two qubit states have the same envelope functions, $\tilde{E}_{HM}$ is effectively independent of $\tilde{\delta}$ (Fig.~2 of Supplement). The sensitivity of the $\sigma_z$ gate to charge noise can be reduced to any desired level by reducing $d\tilde{E}_{HM}/d \tilde{\delta}$.

Charge noise is dominated by the low-frequency part of the spectrum, is long wavelength and \textit{cannot} differentiate valleys. In conventional charge qubits with different charge distributions, charge noise and electron-phonon coupling differentiate the states and lead to dephasing. We have devised a charge qubit that is immune to charge noise, and whose coherence can be further improved through valley-echo. The valley-based qubit states have the same spin configuration, so this architecture is also insensitive to magnetic noise. (A forthcoming publication \cite{Gamble_VlyRlx_11} obtains an intervalley dephasing time of $\mu$s, rather than ns for GaAs charge qubits.) These arguments still hold if the effective mass approximation is relaxed by including ${\bm k}$-states in the vicinity of $\pm {\bm k}_0$.

Charge defects in the dielectric and interface roughness affect the vertical field, and may cause $|\tilde{\Delta}_L| \ne |\tilde{\Delta}_R| $. The operation of the qubit is unaltered, since a \textit{change} in $|\tilde{\Delta}|$ still induces mixing between the eigenstates of $\tilde{H}_T$. Any difference $|\tilde{\Delta}_L| - |\tilde{\Delta}_R|$ can be offset by gate control of $\Delta$ prior to operating the qubit. The proposal is robust with respect to unavoidable differences between the two dots. 

The top gate may alter the phase of $\tilde{\Delta}$ by a small amount \cite{Saraiva_PRB09}, inducing tunneling between the qubit space and the states $\tilde{T}^{LR}_{\pm\pm}$. Yet $\tilde{T}^{LR}_{\pm\pm}$ are separated by $2|\tilde{\Delta}|$, and the intervalley tunneling matrix element is a fraction of the intravalley one (which is $\approx$ tens of $\mu$eV \cite{Culcer_PRB10}, an order of magnitude less than $|\tilde{\Delta}|$ as measured recently \cite{Xiao_Valley_APL10, Borselli_Valley_APL11, Lim_SiQD_SpinFill_NT11}.) 


We have devised a new valley-based QC platform in Si QDs that offers full electrical control with current technology. The qubit is initialized using a Zeeman field, rotated using gate electric fields, and read out via charge sensing. Coupling to a superconducting resonator allows entanglement. The platform is tailored to Si, with the valleys perpendicular to the interface, and $\tilde{\Delta}$ controlled by a top gate \cite{Goswami_NP07, Saraiva_PRB09}. Valley-based qubits in carbon QDs must be studied separately \cite{Churchill_PRL09, Wang_GfnQDwSET_APL10}. Charge-based QC schemes, being easier to control, yield insight into qubit manipulation and coherence, and spur better design.

This work was supported by LPS-NSA-CMTC and by the National Natural Science Foundation of China under Grant No. 91021019. ALS and BK were partially supported by CNPQ, FAPERJ and INCT on Quantum Information. X. H. also thanks NSA/ LPS for support through ARO, DARPA QuEST through AFOSR, and NSF PIF. We thank A. Morello, A. Dzurak, S. Rogge, M. Simmons, A. P\'alyi, H. W. Jiang, R. B. Liu, R. J. Joynt, F. Zwanenburg, N. M. Zimmerman, and G. P. Guo for enlightening discussions.

\section{Supplementary material: One and two-electron states of a multi-valley quantum dot}

Each dot is located at ${\bm R}_D$, with $D = L, R$.  The ground-state wave functions $D_{\xi}$ satisfy $(T + V_D) \, D_\xi = \varepsilon_D \, D_\xi$, with $T$ the kinetic energy operator including the effective mass anisotropy ($m^* \ne m_z$, with $m^*$ the in-plane effective mass as in the text),
\begin{equation}
D_\xi = \phi_D(x,y) \, \psi (z) \, u_\xi ({\bm r})\, e^{ik_\xi z}.
\end{equation}
The lattice-periodic function $ u_\xi ({\bm r}) = \sum_{\bm K} c^\xi_{\bm K} e^{i{\bm K}\cdot{\bm r}}$, with ${\bm K}$ the reciprocal lattice
vectors. The envelopes $\phi_D (x, y)$ are Fock-Darwin states
\begin{equation}
\phi_D (x, y) = \frac{1}{a\sqrt{\pi}} \, e^{-\frac{(x - x_D)^2 + y^2}{2a^2}},
\end{equation}
The EMA equation satisfied by $\psi(z)$ has been given in Refs. \onlinecite{Saraiva_PRB09, Culcer_Roughness_PRB10}.  It can be solved
numerically or analytically. For analytical calculations, a convenient choice is the variational wave function \cite{Bastard},
\begin{equation}
\begin{array}{rl}
\displaystyle \psi(z) = & \displaystyle M \, e^{\frac{k_b z}{2}}, z < 0 \\ [3ex]
= & \displaystyle N\, (z + z_0) \, e^{\frac{-k_{Si} z}{2}}, z > 0,
\end{array}
\end{equation}
where  $k_b = \sqrt{\frac{2m_bU_0}{\hbar^2}}$, and continuity of the wave function at the interface requires $M = Nz_0$.  The only variational
parameter is $k_{Si}$, with the others given in Ref.\ \onlinecite{Culcer_Roughness_PRB10}. The overlap of states from different valleys is
\begin{equation}
\begin{array}{rl}
\displaystyle l_{z, \bar{z}} = & \displaystyle \dbkt{D_z}{D_{\bar{z}}} = \dbkt{\psi(z) e^{ik_0z}}{\psi(z)e^{-ik_0z}} \\ [3ex]
\displaystyle = & \displaystyle \frac{N^2z_0^2}{k_b - 2ik_0} + N^2 \bigg[ \frac{1}{(k_b - 2ik_0)^2} + \frac{z_0}{k_b - 2ik_0} \bigg].
\end{array}
\end{equation}
Since $l_{z, \bar{z}} \ll 1$ by many orders of magnitude we take $D_z$, $D_{\bar{z}}$ to be orthogonal. The overlap is zero in bulk Si, and it
remains zero if the envelopes are wide and smooth compared to the lattice constant. This is the case in our numerical analysis. The appearance
of $l_{z, \bar{z}}$ is an artifact of the trial analytical wave function we have suggested.

The overlap $l = \dbkt{L_\xi}{R_\xi}$ is not zero. It is therefore convenient to construct single-dot wave-functions that are orthogonal, as was done in Ref. \ \onlinecite{Culcer_Roughness_PRB10}, as $\tilde{L}_\xi = \frac{L_\xi - gR_\xi}{\sqrt{1 - 2lg + g^2}}$ and $\tilde{R}_\xi = \frac{R_\xi -
gL_\xi}{\sqrt{1 - 2lg + g^2}}$, where $g = (1 - \sqrt{1 - l^2})/l$, so that $\langle \tilde{R}_\xi | \tilde{L}_\xi \rangle = 0$.  We define
\begin{equation}
\begin{array}{rl}
\displaystyle \tilde{\varepsilon}_0 = & \displaystyle \tbkt{\tilde{D}_\xi}{T + V_D}{\tilde{D}_\xi} \\ [3ex]
\displaystyle \tilde{\Delta} = & \displaystyle \tbkt{\tilde{D}_\xi}{U_0 \theta(z) + eFz}{\tilde{D}_{-\xi}} \\ [3ex]
\displaystyle \tilde{t} = & \displaystyle \tbkt{\tilde{L}_\xi}{H_0}{\tilde{R}_\xi} + \tbkt{\tilde{L}_\xi \tilde{L}_\xi}{V_{ee}}{\tilde{L}_\xi
\tilde{R}_\xi} \equiv \tilde{t}_0 + \tilde{s}.
\end{array}
\end{equation}
We subsequently orthogonalize $L_\pm$ and $R_\pm$ obtaining $\tilde{L}_\pm = \frac{L_\pm - gR_\pm}{\sqrt{1 - 2lg + g^2}}$ and $\tilde{R}_\pm =
\frac{R_\pm - gL_\pm}{\sqrt{1 - 2lg + g^2}}$. These are the states used throughout this work.

\begin{figure}[tbp]
\includegraphics[width=\columnwidth]{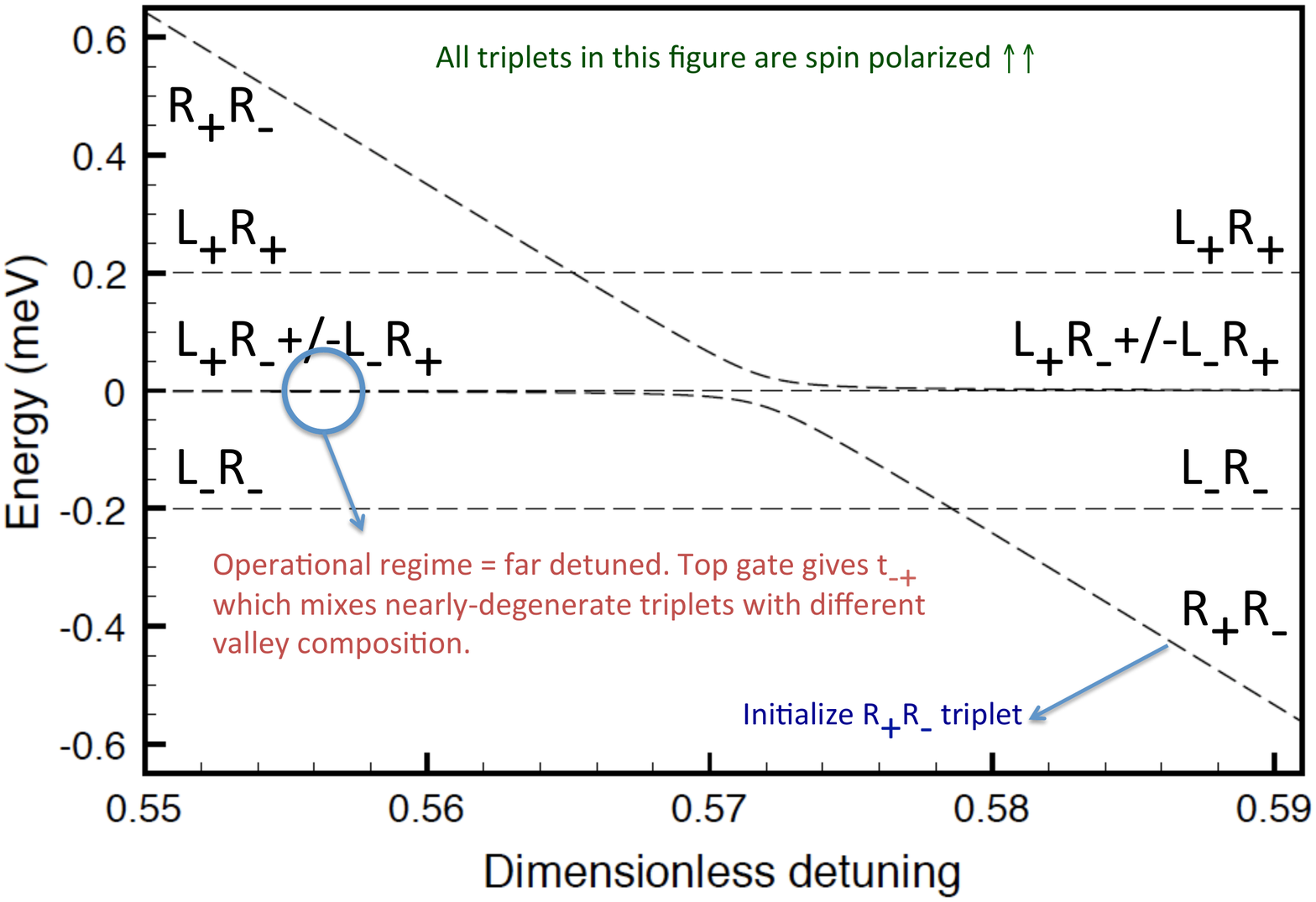}
\caption{Spin-polarized $\downarrow\downarrow$ triplet levels in a DQD, as determined in Ref.~\onlinecite{Culcer_Roughness_PRB10}, with a valley splitting of 0.1meV.} \label{fig:levels}
\end{figure}

Two electrons are initialized on one dot. The four lowest-energy two-particle spatial wave functions are,
\begin{equation}\label{SingleDot}
\arraycolsep 0.3 ex
\begin{array}{rl}
\displaystyle \tilde{S}^{RR}_{\pm \pm} = & \displaystyle \tilde{R}_\pm^{(1)} \tilde{R}_\pm^{(2)}
\\ [1ex]
\displaystyle \tilde{S}^{RR}_{+-} = & \displaystyle (1/\sqrt{2}) \big(\tilde{R}_+^{(1)} \tilde{R}_-^{(2)}  + \tilde{R}_+^{(2)}
\tilde{R}_-^{(1)}\big)
\\ [1ex]
\displaystyle \tilde{T}^{RR}_{+-} = & \displaystyle(1/\sqrt{2}) \big(\tilde{R}_+^{(1)} \tilde{R}_-^{(2)} - \tilde{R}_+^{(2)}
\tilde{R}_-^{(1)}\big),
\end{array}
\end{equation}
with the superscript $i$ denoting particle $i$. The following two-electron matrix elements are needed
\begin{equation}
\arraycolsep 0.3 ex
\begin{array}{rl}
\displaystyle \tilde{k} = & \displaystyle \tbkt{\tilde{L}_\xi^{(1)}
\tilde{R}_\xi^{(2)}}{V_{ee}}{\tilde{L}_\xi^{(1)}  \tilde{R}_\xi^{(2)}} = \tbkt{\tilde{L}_\xi^{(1)} \tilde{R}_{-\xi}^{(2)}}{V_{ee}}{\tilde{L}_\xi^{(1)}
\tilde{R}_{-\xi}^{(2)}}  \\ [1ex] \displaystyle
\tilde{j} = & \displaystyle \tbkt{\tilde{L}_\xi^{(1)}
\tilde{R}_\xi^{(2)}}{V_{ee}}{\tilde{L}_\xi^{(2)}  \tilde{R}_\xi^{(1)}} = \tbkt{\tilde{L}_\xi^{(1)}
\tilde{R}_{-\xi}^{(2)}}{V_{ee}}{\tilde{L}_{-\xi}^{(2)}  \tilde{R}_\xi^{(1)}} \\ [1ex] \displaystyle \tilde{s} =
& \displaystyle \tbkt{\tilde{L}_\xi^{(1)}
\tilde{L}_\xi^{(2)}}{V_{ee}}{\tilde{L}_\xi^{(1)}  \tilde{R}_\xi^{(2)}} = \tbkt{\tilde{L}_\xi^{(1)}
\tilde{L}_{-\xi}^{(2)}}{V_{ee}}{\tilde{L}_\xi^{(1)}  \tilde{R}_{-\xi}^{(2)}} \\ [1ex] \displaystyle
\tilde{u} = & \displaystyle \tbkt{\tilde{D}_\xi^{(1)}  \tilde{D}_\xi^{(2)}}{V_{ee}}{\tilde{D}_\xi^{(1)}
\tilde{D}_\xi^{(2)}} = \tbkt{\tilde{D}_\xi^{(1)}  \tilde{D}_{-\xi}^{(2)}}{V_{ee}}{\tilde{D}_\xi^{(1)}
\tilde{D}_{-\xi}^{(2)}}.
\end{array}
\end{equation}
In the basis $\{ \tilde{S}^{RR}_{- -}, \tilde{S}^{RR}_{+ -}, \tilde{T}^{RR}_{+ -}, \tilde{S}^{RR}_{++} \}$ the Hamiltonian is
\begin{equation}\label{1dot}
H_{1dot} = 2 \tilde{\varepsilon}_0 + \tilde{u} + \begin{pmatrix}
-2|\tilde{\Delta}| & 0 & 0 & 0 \cr
0 & 0 & 0 & 0 \cr
0 & 0 & 0 & 0 \cr
0 & 0& 0 & 2|\tilde{\Delta}|
\end{pmatrix}.
\end{equation}
A Zeeman field can separate the polarized triplets, and for a Zeeman energy $E_Z > 2\Delta$, the lowest level is the polarized triplet $\tilde{T}^{RR, \downarrow\downarrow}_{+-}$.

\begin{figure}[tbp]
\includegraphics[width=\columnwidth]{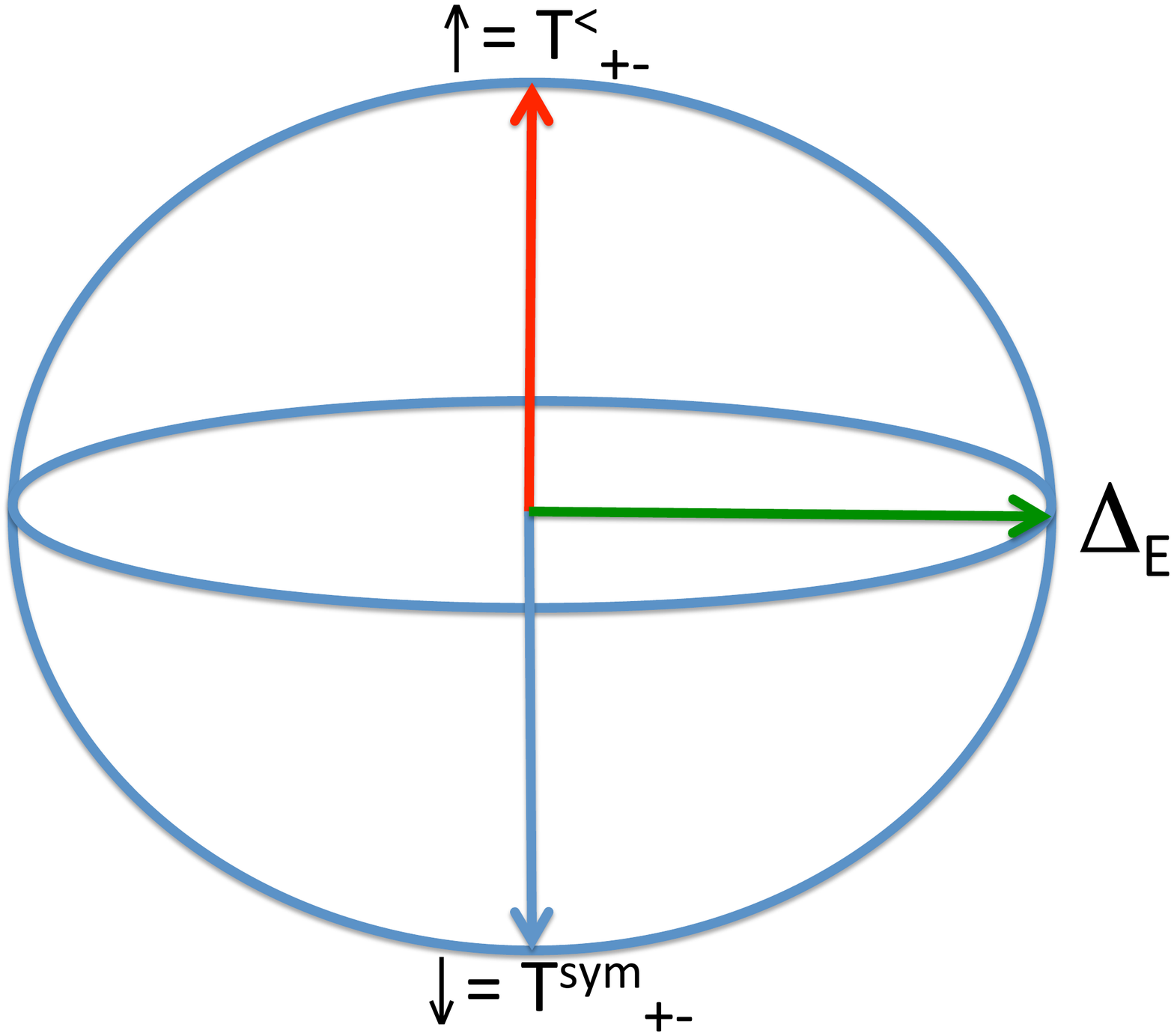}
\caption{Bloch sphere representation of the qubit states $\tilde{T}^<_{+-}$ and $\tilde{T}^{sym}_{+-}$. The $\sigma_x$ gate is provided by the top-gate induced difference $\tilde{\Delta}_E$ in the valley-orbit coupling between the two dots. The $\sigma_z$ gate (not shown) is provided by the detuning.}
\label{fig:Bloch}
\end{figure}

When both dots are considered, we will only require two-electron spin-triplet states. For an asymmetric DQD, in which the potential of the left dot is raised, we may neglect the $\tilde{L}\tilde{L}$ states. The remaining triplets are
\begin{equation}
\arraycolsep 0.3 ex
\begin{array}{rl}
\displaystyle \tilde{T}^{LR}_{\pm\pm} = & \displaystyle (1/\sqrt{2}) \, \big( \tilde{L}_\pm^{(1)} \tilde{R}_\pm^{(2)} - \tilde{L}_\pm^{(2)}
\tilde{R}_\pm^{(1)} \big) \\ [1ex]
\displaystyle \tilde{T}^{LR}_{\pm\mp} = & \displaystyle (1/\sqrt{2}) \, \big( \tilde{L}_\pm^{(1)} \tilde{R}_\mp^{(2)} - \tilde{L}_\pm^{(2)}
\tilde{R}_\mp^{(1)} \big)
\\ [1ex]
\displaystyle \tilde{T}^{RR}_{+-} = & \displaystyle (1/\sqrt{2}) \, \big( \tilde{R}_+^{(1)} \tilde{R}_-^{(2)} - \tilde{R}_+^{(2)}
\tilde{R}_-^{(1)} \big),
\end{array}
\end{equation}

In this work we only consider the polarized $\downarrow\downarrow$ triplets, thus in all of the above it is understood that e.g.
$\tilde{T}^{LR}_{\pm\pm} \equiv \tilde{T}^{LR, \downarrow\downarrow}_{\pm\pm}$, and so forth. The effective Hamiltonian in the basis $\{\tilde{T}^{LR}_{- -}, \tilde{T}^{LR}_{++}, \tilde{T}^{LR}_{+-}, \tilde{T}^{LR}_{-+}, \tilde{T}^{RR}_{+-}\}$ is
\begin{equation}\label{H_T_long}
\arraycolsep 0.3 ex
\begin{array}{rl}
\displaystyle \tilde{H}_T = & \displaystyle \begin{pmatrix}
- |\tilde{\Delta}|_{tot} - \tilde{j} & 0 & 0 & 0 & 0 \cr
0 & |\tilde{\Delta}|_{tot} - \tilde{j} & 0 & 0 & 0 \cr
0 & 0 & |\tilde{\Delta}|_E & -\tilde{j} & \tilde{t} \cr
0 & 0 &  -\tilde{j} & - |\tilde{\Delta}|_E & - \tilde{t} \cr
0 & 0 & \tilde{t} & - \tilde{t} & - \tilde{\delta}
\end{pmatrix},
\end{array}
\end{equation}
When the Hamiltonian $H_T$ of Eq.\ (\ref{H_T_long}) is diagonalized for $\tilde{t}_- = 0$, the eigenstates are $\tilde{T}^{LR}_{\pm\pm}$ and
\begin{equation}
\begin{array}{rl}
\displaystyle \tilde{T}^>_{+-} = & \displaystyle \frac{\varepsilon^>_0}{\sqrt{\varepsilon^{>2}_0 + 2\tilde{t}^2}} \,
\bigg( \frac{\tilde{t}\sqrt{2}}{\varepsilon^>_0} \, \tilde{T}^{anti}_{+-} + \tilde{T}^{RR}_{+-} \bigg) \\ [3ex]
\displaystyle \tilde{T}^<_{+-} = & \displaystyle \frac{\varepsilon^<_0}{\sqrt{\varepsilon^{<2}_0 + 2\tilde{t}^2}} \,
\bigg( \frac{\tilde{t}\sqrt{2}}{\varepsilon^<_0} \, \tilde{T}^{anti}_{+-} + \tilde{T}^{RR}_{+-} \bigg) \\ [3ex]
\displaystyle \tilde{T}^{sym}_{+-} = & \displaystyle \frac{1}{\sqrt{2}} \, ( \tilde{T}^{LR}_{+-} + \tilde{T}^{LR}_{-+}) \\ [3ex]
\displaystyle \tilde{T}^{anti}_{+-} = & \displaystyle \frac{1}{\sqrt{2}} \, (\tilde{T}^{LR}_{+-} - \tilde{T}^{LR}_{-+}).
\end{array}
\end{equation}
The energies are
\begin{equation}
\begin{array}{rl}
\displaystyle \varepsilon^{>}_0 = & \displaystyle \frac{ -\tilde{\delta} + \tilde{j} + \sqrt{(\tilde{\delta} + \tilde{j})^2 +
8\tilde{t}^2} }{2} \\ [3ex]
\displaystyle \varepsilon^{<}_0 = & \displaystyle \frac{ -\tilde{\delta} + \tilde{j} - \sqrt{(\tilde{\delta} + \tilde{j})^2 +
8\tilde{t}^2} }{2}.
\end{array}
\end{equation}
Figure~\ref{fig:levels} shows the five energy levels as a function of the dimensionless detuning in the case of no valley coupling difference across the two dots. The qubit is defined by the ground and second excited states on the right of Fig.~\ref{fig:levels}, with energies $\tilde{\varepsilon}_0^<$ and 0. Figure~\ref{fig:Bloch} illustrates the qubit explicitly on the Bloch sphere.

Valley-echo removes the effects of inhomogeneous broadening and the associated dephasing. The state $\tilde{T}^{RR}_{+-}$ is initialized and swept to $\tilde{T}^{<, \uparrow\uparrow}_{+-}$, then $\tilde{\Delta}_E$ is set to a fixed value, controlling the $\sigma_x$ rotation. After a specified time $\tau$ the detuning is swept so as to apply a fast $\pi$-pulse of the $\sigma_z$ gate. Readout is performed at time $2\tau$, when the qubit returns to its initial state.

\end{document}